# Electrically modulated dynamic spreading of drops on soft surfaces

Ranabir Dey[1], Ashish Daga[1], Sunando DasGupta[2,3], Suman Chakraborty[1,3*]

[1]*Department of Mechanical Engineering, Indian Institute of Technology Kharagpur, Kharagpur- 721 302, West Bengal, India.*

[2]*Department of Chemical Engineering, Indian Institute of Technology Kharagpur, Kharagpur- 721 302, West Bengal, India.*

[3]*Advanced Technology Development Centre, Indian Institute of Technology Kharagpur, Kharagpur- 721 302, West Bengal, India.*

## ABSTRACT

The intricate interaction between the deformability of a substrate and the dynamic spreading of a liquid drop on the same, under the application of an electrical voltage, has remained far from being well understood. Here, we demonstrate that electrospreading dynamics on soft substrates is dictated by the combined interplay of electrocapillarity, the wetting line friction and the viscoelastic energy dissipation at the contact line. Our results reveal that during such electro-elastocapillarity mediated spreading of a sessile drop, the contact radius evolution exhibits a universal power law in a substrate elasticity based non-dimensional time, with an electric potential dependent spreading exponent. Simultaneously, the macroscopic dynamic contact angle variation follows a general power law in the contact line velocity, normalized by elasticity dependent characteristic velocity scale. Our results are likely to provide the foundation for the development of a plethora of new applications involving droplet manipulations by exploiting the interplay between electrically triggered spreading and substrate-compliance over interfacial scales.

---

[*] Corresponding author, E-mail: suman@mech.iitkgp.ernet.in

Rapid advancements in miniaturized electronics and biomedical sciences have triggered a wide range of scientific investigations, demanding precise control over the manipulation of tiny volumes of droplets in engineered devices [1–3]. This has eventually lead to droplets being intricately manoeuvred by exploiting the electrically tunable gradients of interfacial tension [4–6]. Such electrical control over droplet wetting and subsequent dewetting dynamics has recently been translated into several novel applications such as variable focus-lenses [5,6], electronic displays/e-papers [5,6], droplet manipulation over functional substrates [7–9], control of interfaces between immiscible solutions [10], thermal management of miniaturized devices [11], and biochemical detection [12], to name a few. However, despite such remarkable advancements from a technological perspective, the physics of electrically-induced dynamic spreading of liquid droplets on soft substrates remains far from being well understood. This deficit stems from the complexities in addressing the non-trivial interplay between electrically induced modulation of capillary forces and substrate deformation induced interfacial interactions over the relevant spatio-temporal scales.

In the classical literature, electrospreading of sessile droplets on solid substrates has been commonly addressed without taking the substrate deformability in purview. Fundamentally, the spontaneous spreading of a sessile droplet, on the application of an electric potential, $V$, is initiated by the electrostatic reduction of the solid-liquid interfacial energy between the droplet and the insulating surface [4–6] [Fig. 1(a)]. During such electrospreading process on non-deformable substrates, the spreading dynamics, in the regime of high contact line velocity $(v_{cl})$, is predominantly controlled by the energy dissipation due to the three-phase contact line (TPCL) friction [13-17]. On the other hand, the variation in the macroscopic dynamic contact angle ($\theta_d$), for lower values of $v_{cl}$, conforms to a hydrodynamic description, stemming from the matching of the internal Stokes flow solution in the quasi-static macroscopic region to that in the mesoscopic viscocapillary region of the electrically-mediated advancing droplet meniscus [15, 17-20]. However, on immediate contact between the droplet and the dielectric surface under an applied electric field, the wetting dynamics is primarily dictated by the balance between capillary and inertial forces [21, 22], unveiling a temporal regime that falls beyond the focus of the present report.

The aforementioned descriptions of electrospreading of sessile droplets are intrinsically restricted to rigid substrates. On the contrary, a sessile droplet resting on a soft substrate triggers a spontaneous substrate deformation [23-27]. The characteristic length-scale for this substrate deformation is $\gamma/E$, where $\gamma$ is the liquid surface tension and $E$ is the Young's modulus of the substrate [23-26, 28]. In this respect, it is well established that such microscopic deformation of the soft surface alters the macroscopic spontaneous spreading/dewetting dynamics of liquid drops on the same [29-35]. However, the inter-connection between electrically-induced spreading of a drop and the deformation of the supporting, soft substrate still remains unclear.

Here, we bring out the temporal variations in macroscopic dynamic contact angle ($\theta_d$) and contact radius ($r_c$) of millimetre-sized sessile drops during their electrically modulated spreading on rheologically-tunable soft dielectric films, for different magnitudes of the applied electric potential ($V$). Unlike previous reports on electrowetting and electrospreading, our analysis takes into account the deformability of the dielectric substrate, instead of treating the same as trivially rigid. Our results reveal that the electrospreading dynamics on soft, deformable surfaces is controlled by the combined influences of electrocapillarity, the wetting line friction and the viscoelasticity of the substrate material. We further show that on the soft substrate, the normalized $r_c$ evolves with a dimensionless time (normalized based on the substrate elasticity) in a universal power-law, where the spreading exponent increases with increasing $V$. In addition, we demonstrate that $\theta_d$ varies with normalized $v_{cl}$ according to another universal power law, where $v_{cl}$ is non-dimenisonalized by the elasticity dependent characteristic velocity scale. We contemplate our experimental results with suitable scaling arguments, by appealing to a dynamic energy conservation based model.

Dielectric films of varying elasticity are prepared from Sylgard 184 (Dow Corning, USA; dielectric constant, $\varepsilon_r = 2.65$) - a Polydimethylsiloxane (PDMS) based elastomer. The prepolymer is prepared by mixing the base and the curing agent in the weight ratios of 10:1, 30:1 and 50:1. The degassed prepolymer is then spin-coated on transparent Indium Tin Oxide (ITO) coated glass slides, followed by subsequent overnight curing at 95 °C. According to reported trends, bulk rheometry tests reveal that $E$ of the differently cross-linked Sylgard 184 films decreases from 1.5 MPa to 0.02 MPa, with increasing base-to-cross-linker ratio [34, 35]. Moreover, the different Sylgard 184 films are also characterised by evaluating the equilibrium contact angle ($\theta_{eq}^0$), the macroscopic advancing and receding contact angles

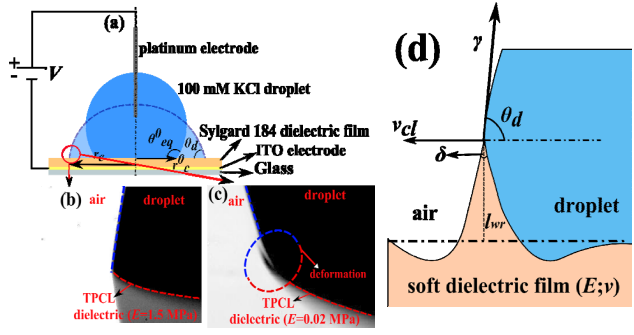

FIG.1 (a) Schematic of the droplet-and-dielectric system used for investigating the electrospreading behaviour of sessile drops. The detailed experimental setup is shown in [36; Fig. S1]. (b) Magnified images of the advancing droplet meniscus, near the TPCL, on the rigid dielectric film ($E = 1.5$ MPa), and (c) on the soft dielectric film ($E = 0.02$ MPa), at 160 V. For $E = 1.5$ MPa, the droplet and the dielectric film form a sharp solid-liquid-air interface (TPCL; red dashed line). However, for $E = 0.02$ MPa, there is an apparent deformation of the film surface about the TPCL. (d) Schematic showing capillarity-induced deformation of the soft film about the TPCL. The localized modification of $\theta_d$ due to the deformation is neglected, on a macroscopic scale, for millimetre-sized droplets [24, 29-33, 35].

($\theta_a^0 / \theta_r^0$), and the initial (equilibrium) contact radius ($r_c^0$) of aqueous 100 mM KCl solution drops on the films, without considering any electrical effects [36; Table S1]. During the experiments, a sessile droplet (volume: $5 \pm 1\,\mu l$) of aqueous 100 mM KCl solution is first dispensed onto the dielectric layer of thickness $h$, by a calibrated microsyringe. The droplet spreading, initiated on application of the DC electric potential ($V$) between the platinum wire electrode, immersed in the sessile droplet, and the ITO electrode [Fig. 1(a)], is recorded at 2000 fps by a high speed camera fitted with a macro zoom lens. In a further attempt to qualitatively resolve the deformation of the soft substrate surfaces about the TPCL, during the electrospreading process, high speed magnified images are also recorded by using a microscopy lens [Fig. 1(b), Fig. 1(c) and Fig. 1(d)]. For the different dielectric films, identical experiments are performed at different voltages, between 100 V and 160 V. The resulting variations in $\theta_d$ and $r_c$ are evaluated from subsequent image processing by using ImageJ software [37; for image processing details see 36 (Fig. S4)].

The temporal variations of $\theta_d$ and $r_c$ during the electrospreading process, on the dielectrics of varying elasticity, are shown in Fig. 2(a) and Fig. 2(b) respectively. Here, $\bar{\theta}_d = \theta_d / \theta_{eq}^0$ and $\bar{r}_c = r_c / r_c^0$. On application of the electric potential, the rate of decrease of $\bar{\theta}_d$ is distinctly faster on the substrate with greater elasticity [Fig. 2(a)]. However, it is the variation of $\bar{r}_c$ that clearly indicates that, for a particular magnitude of $V$, the electrospreading rate appreciably decreases, and hence, the total spreading time consequently increases, with decreasing $E$ [Fig. 2(b)]. Furthermore, for a particular magnitude of $V$, the final steady-state value of $r_c$ progressively decreases, and the corresponding measured equilibrium contact angle value increases, with decreasing $E$ [Fig 2(a) and Fig. 2(b)]. To summarize the qualitative scenario, the rate and extent of electrospreading, for a particular magnitude of $V$, decrease with increasing softness of the insulating surface. Additionally, on a particular dielectric film, the same quantities increase with increasing $V$ [Fig. 2(a) and Fig. 2(b)].

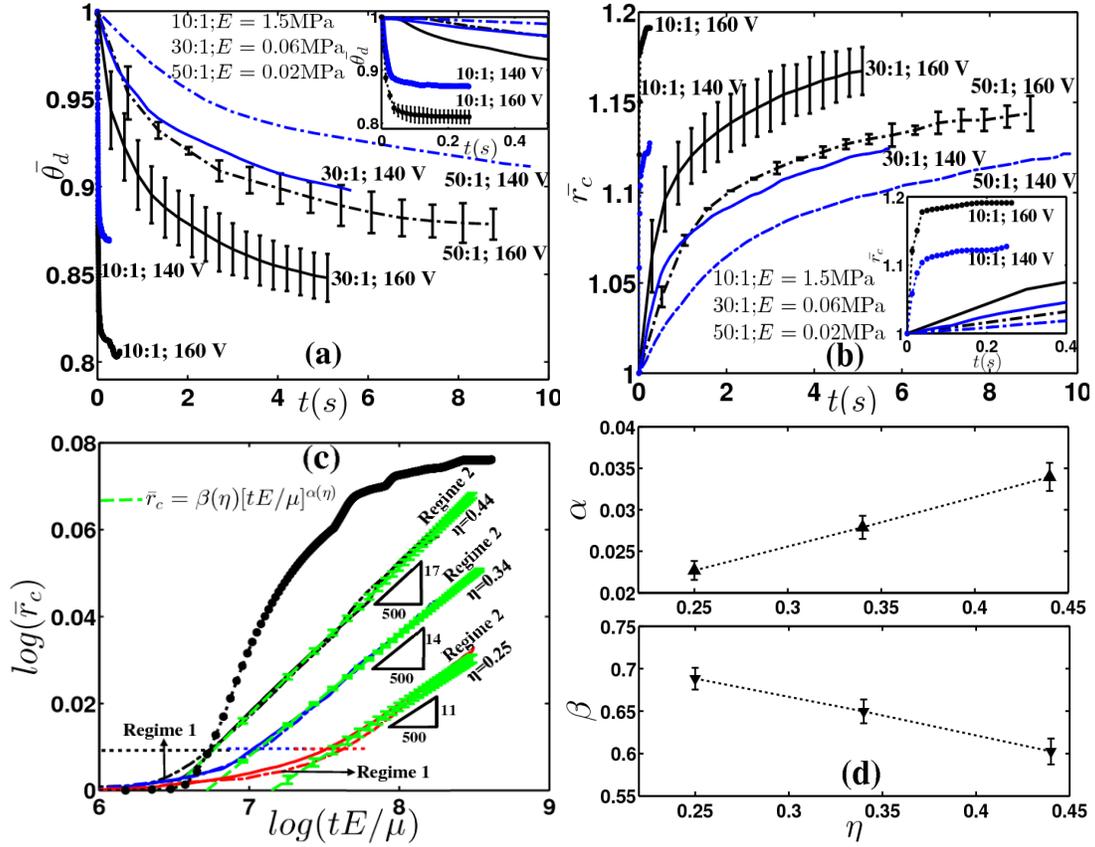

FIG. 2 Temporal variations of (a) the non-dimensional macroscopic dynamic contact angle ($\bar{\theta}_d$), and (b) the non-dimensional contact radius ($\bar{r}_c$), during electrospreading on dielectric films of varying elasticity, at 140 V (blue) and 160 V (black). The experimental data shown here represent the average values obtained after repeated trials. The errorbars are shown only for the data corresponding to 160 V to maintain clarity. (c) Collapse of the $\bar{r}_c$ vs. $t$ data sets for the soft dielectric films into distinct master curves, one corresponding to each value of $\eta$, on rescaling $t$ with a substrate elasticity dependent characteristic time scale ($\mu/E$). The deviation of the spreading characteristics on the rigid film ($E$=1.5 MPa), from the universal behavior on the soft films ($O(\gamma/Er_c^0) \geq 10^{-3}$), is shown only for $\eta = 0.44$, to maintain clarity. The electrospreading (*Regime 2*) master curves follow a power-law in $tE/\mu$ (shown here with the dashed line [– – –], along with the standard error of the estimate), where the prefactor ($\beta$) and the power-law exponent ($\alpha$) are functions of $\eta$, as shown in (d).

Next, by executing dimensional analysis, we evaluate the evolution of $\bar{r}_c$ in the following functional form: $\bar{r}_c = f\left(\dfrac{t}{\mu/E};\eta\right)$, where $\mu$ is the liquid viscosity and $\eta = \varepsilon_r \varepsilon_0 V^2 / 2h\gamma$ is a non-dimensional parameter representing the ratio of the electrostatic reduction in the specific solid-liquid interfacial energy to the liquid surface tension, corresponding to $V$ [5]. On non-dimensionalizing the $\bar{r}_c$ vs. $t$ data sets, as shown in Fig. 2(b), with the particular time scale $\mu/E$, the data sets for the soft dielectric films (i.e. for $E$=0.06 MPa and $E$=0.02 MPa) collapse into distinct master curves, each corresponding to a unique value of $\eta$ [Fig. 2(c)]; while the electrospreading characteristics for the rigid film ($E$=1.5 MPa) deviate from these master curves. Hence, it can be now concluded that the droplet electrospreading dynamics, on a soft solid surface, is intrinsically controlled by the combined influences of the elasticity of the solid and the liquid viscosity [38], and is not generally independent of the rheological properties of the dielectric, as portrayed in the existing literature [5, 13-17, 39]. Furthermore, this characteristic time scale can be recast as $\mu/E = \left(\mu r_c^0 / \gamma\right) \times \left(\gamma / E r_c^0\right)$, to include the factor, $\dfrac{\gamma/E}{r_c^0}$, which represents the non-dimensional length scale for the elastocapillarity-induced deformation of the soft substrates [23-26, 28]. This further highlights the fact that the electrospreading dynamics on soft solids ($O\left(\gamma/Er_c^0\right) \geq 10^{-3}$) is dependent on the elastocapillarity of the droplet and dielectric system. Moreover, the decrease in the electrospreading rate, with decreasing $E$, can be physically attributed to the increasing substrate deformation on the softer solids.

The log-log plot in Fig. 2(c) also reveals that the master curves exhibit two distinct regimes. The regime attained over early dimensionless times is marked as *Regime 1*, over which no substantial change in $\bar{r}_c$ occurs. On the other hand, in *Regime 2*, $\bar{r}_c$ has the power-law form: $\bar{r}_c = \beta(\eta)[tE/\mu]^{\alpha(\eta)}$. Here, the prefactor $\beta(\eta)$ and the power-law exponent $\alpha(\eta)$ are functions of $\eta$ [Fig. 2(d)]. Hence, Fig. 2(c) and Fig. 2(d) clearly indicate that the temporal evolution of $\bar{r}_c$ during electrospreading on soft films, towards the new equilibrium state, conforms to a universal power-law in a non-dimensional time ($tE/\mu$), where the

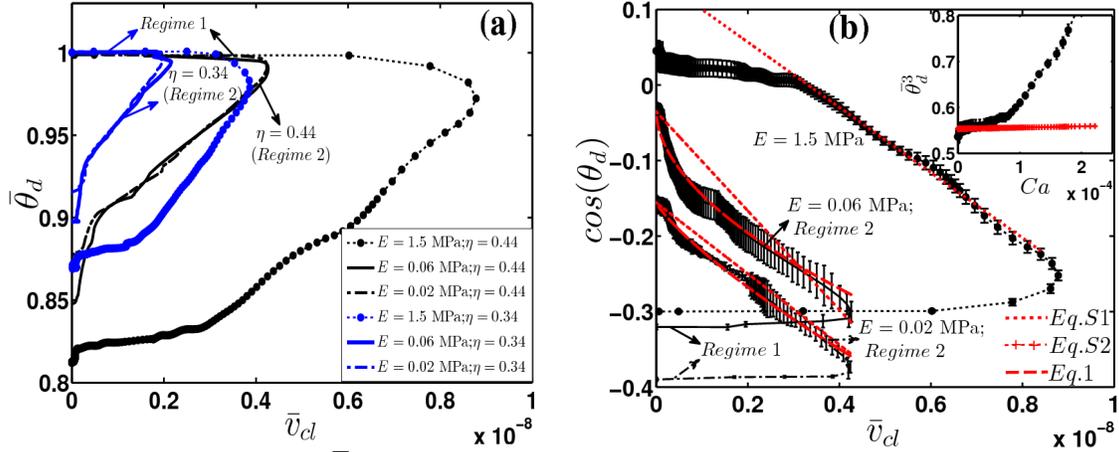

FIG. 3 (a) Collapse of the $\bar{\theta}_d$ vs. $v_{cl}$ data sets, for the soft dielectric films, into master curves, one for a definite value of $\eta$, on reducing $v_{cl}$ with a substrate elasticity dependent characteristic velocity scale ($r_c^0 E/\mu$). The experimental data shown here represent the average values obtained after repeated trials. (b) Description of the variation of $\theta_d$ with $\bar{v}_{cl}$ through a dynamic energy balance approach (shown here for $\eta = 0.44$). The electrospreading dynamics on the rigid film ($E$=1.5 MPa) adheres to the classical descriptions given by Eq. S1 [36], in the range of higher values of $v_{cl}$, and Eq. S2 [36], in the range of lower values of $v_{cl}$ (see inset). However, the variation of $\theta_d$ with $\bar{v}_{cl}$, for the soft films, is suitably addressed by the dynamic energy balance represented by Eq. 1, and not by Eq. S1 or Eq. S2. For the physically consistent values (obtained by a nonlinear least-squares fitting method) of the parameters involved, see [36].

power-law index (i.e. the spreading exponent) increases with increasing $\eta$.

On normalizing $v_{cl}$ with a characteristic velocity scale, $v_o = (r_c^0 E)/\mu$, commensurate with the characteristic time scale $\mu/E$, $\bar{\theta}_d$ variations with $v_{cl}$ on the soft substrates also collapse into master curves, one for a definite value of $\eta$ [Fig. 3(a)]. Fig. 3(a) further accentuates the dependence of the electrically-induced contact line velocity on the degree of compliance of the dielectric film surface, under the forcing imposed by the moving TPCL. The depicted characteristics also indicate that while $v_{cl}$ is determined by both $\eta$ and $E$, $\bar{v}_{cl}(\eta) = v_{cl}/v_o = (\mu v_{cl})/(r_c^0 E)$, on soft solids, is dependent only on $\eta$.

To physically comprehend the aforementioned dependence of the electrospreading dynamics on the substrate elasticity, we resort to a dynamic energy balance model. To highlight the efficacy of this model, we use the electrospreading behaviour on the film with $E = 1.5$ MPa (rigid substrate) as a control study [Fig. 3(b)]. One may note here that in sharp contrast to the classical mathematical description of electrospreading on rigid substrates (see

Eq. S1, S2 in [36]), the uniqueness of the electrospreading dynamics on the soft substrates stems from the involved motion of the elastocapillarity-induced wetting ridge [Fig. 1(c) and Fig. 1(d)] on the soft dielectric surface, along with the advancing TPCL. The associated rate of expenditure of energy, per unit length of the TPCL, in 'pulling up' this wetting ridge by the liquid surface tension, during axisymmetric droplet spreading is given as [29-33, 35]: $\dot{W} \approx \frac{2\gamma^2(1-\nu^2)}{\pi E \delta} v_{cl} (\sim (l_{wr} \gamma v_{cl})/r_c$, where $l_{wr} \sim \gamma/E$), following the linear elasticity theory. Here, $\nu$ is the Poisson's ratio ($\nu = 0.5$ for incompressible materials), and $\delta$ is the (sub-nanometric) radius of the region, about the TPCL, within which the linear elasticity theory is inapplicable. Due to the intrinsic viscoelastic nature of the soft solids (particularly elastomers), a rate-dependent fraction, $\Delta = (v_{cl}/v_c)^n$, of the stored strain energy ($\dot{W}$) is dissipated, while the rest is restituted, as the soft solid relaxes after the passage of the TPCL and the concomitant displacement of the wetting ridge to a new position [30-33]. Here, the characteristic velocity, $v_c \sim v_o$, and $n$ is a power-law index indicative of the surface damping properties. Hence, on the soft films, as the droplet evolves towards its new thermodynamic equilibrium state due to the electrically-induced imbalance in capillarity, the rate of dissipation of the excess free energy is given by: $\dot{F} = \gamma \left[ \cos\theta_{eq}(\eta;E) - \cos\theta_d(t) \right] v_{cl} \approx \zeta v_{cl}^2 + \dot{W}\Delta$, where $\zeta v_{cl}^2$ is the inherent contact line dissipation ($\zeta$ is the co-efficient of wetting line friction; [18, 19]); while $\dot{W}\Delta$ represents the aforementioned viscoelastic dissipation. On rearrangement, this dynamic energy balance can be written in a non-dimensional form as:

$$\cos\theta_d \approx \cos\theta_{eq}(\eta;E) - \frac{\zeta v_o}{\gamma} \bar{v}_{cl}(\eta) - \frac{3\gamma}{2\pi E \delta} \left[ \bar{v}_{cl}(\eta) \right]^n \quad (1)$$

Eq. 1 (and not Eq. S1 or Eq. S2) captures the general spreading behaviour [40] on soft substrates ($O(\gamma/Er_c^0) \geq 10^{-3}$), as dictated by the electrocapillarity and elastocapillarity of the droplet and dielectric system, for varying magnitudes of $\eta$ [Fig. 3(b)]. Here, $\theta_{eq}(\eta;E)$ specifies the final macroscopic equilibrium condition as uniquely determined by the combined influences of the electric field ($\eta$) and substrate elasticity ($E$) ([36]; Table S2). On the other hand, the non-dimensional co-efficients- $(\zeta v_0)/\gamma$ and $(3\gamma)/(2\pi E \delta)$, and $n$ are constants independent of $\eta$ and $E$ [36; Fig. S2]. The constancies of the non-dimensional co-efficients, over the range of $E$ and $\eta$, are maintained due to the fact that $\zeta$ and $\delta$ increase with decreasing $E$, while simultaneously being independent of $V$ [36; Fig. S3]. This spreading regime, dictated by Eq. 1, constitutes the 'electrospreading paradigm' on soft dielectrics, identified previously as *Regime 2* [Fig. 3(b)]; while it can be now unambiguously stated that *Regime 1* simply represents the regime over which $v_{cl}$ almost instantaneously accelerates to its maximum magnitude, with no substantial change in $r_c$ [Fig. 2(c)] and no significant reduction in $\theta_d$ [Fig. 3(a) and Fig. 3(b)], on application of $V$. *Regime 1* is usually never accounted for in the electrospreading analysis, and the maximum velocity is considered to occur at $t = 0$ [13-17].

To summarize, we have attempted to bring out the effects of substrate deformation on electrospreading of liquid droplets. Our studies have revealed that the additional viscoelastic dissipation acts as one of the primary rate-controlling factors for the electrospreading process on soft dielectrics, in addition to the strength of the electrical actuation. During such electro-elastocapillarity-mediated spreading, the contact radius evolution follows a universal, power-law in substrate elasticity based non-dimensional time. Additionally, the reduction in the macroscopic dynamic contact angle adheres to a general power-law function of the non-dimensional contact line velocity, consistent with a dynamic energy balance. These findings are likely to open up scopes for new technological developments, such as soft flexible lenses, flexible electronic displays for better configurability and advanced digital microfluidic devices for droplet manipulations. The new understanding may also be used to design applications involving soft biological interfaces-like control of spreading in bio-physical processes, as well as for designing systems for controlled drug delivery.

---